\documentclass[journal]{IEEEtran}
\usepackage{multirow}
\usepackage{graphicx}
\usepackage{subfigure}
\usepackage{makecell}
\usepackage{hyperref}
\usepackage{threeparttable}
\usepackage{booktabs}
\usepackage{color}
\usepackage{amsmath}
\usepackage{amssymb}
\usepackage[justification=centering]{caption}
\usepackage{url}

\ifCLASSOPTIONcompsoc
  \usepackage[nocompress]{cite}
\else
  \usepackage{cite}
\fi
\ifCLASSINFOpdf
\else
\fi
\usepackage{ragged2e} 

\begin{document}
%

\title{The QXS-SAROPT Dataset for Deep Learning in SAR-Optical Data Fusion}

\author{Meiyu Huang, Yao Xu, Lixin Qian, Weili Shi, Yaqin Zhang, \\Wei Bao, Nan Wang, Xuejiao Liu, Xueshuang Xiang
	\thanks{The authors are with Qian Xuesen Laboratory of Space Technology, China Academy of Space Technology, Beijing, China. (E-mail: huangmeiyu@qxslab.cn; xuyao@qxslab.cn; qianlixin@whu.edu.cn; shiweili@qxslab.cn; zhangyaqin202102@163.com; baowei97@163.com; wnan2020@foxmail.com; liuxuejiao@qxslab.cn; xiangxueshuang@qxslab.cn)}
	\thanks{Meiyu Huang, Yao Xu and Lixin Qian contributed equally to this work.}
    \thanks{Corresponding author: Xueshuang Xiang.}
\thanks{
	This work is supported by the Beijing Nova Program of Science and Technology under Grant Z191100001119129 and the National Natural Science Foundation of
	China 61702520.}
}
%
%

\markboth{Journal of \LaTeX\ Class Files,~Vol.~14, No.~8, August~2015}%
{Shell \MakeLowercase{\textit{et al.}}: Bare Demo of IEEEtran.cls for IEEE Journals}
%




\IEEEtitleabstractindextext{%
\begin{abstract}
\justifying\let\raggedright\justifying
Deep learning techniques have made an increasing impact on the field of remote sensing. However, deep neural networks based fusion of multimodal data from different remote sensors with heterogenous characteristics has not been fully explored, due to the lack of availability of big amounts of perfectly aligned multi-sensor image data with diverse scenes of high resolutions, especially for synthetic aperture radar (SAR) data and optical imagery. 
To promote the development of deep learning based SAR-optical fusion approaches, we release the QXS-SAROPT dataset, 
which contains 20,000 pairs of SAR-optical image patches. 
We obtain the SAR patches from SAR satellite GaoFen-3 images and the optical patches from Google Earth images.
These images cover three port cities: San Diego, Shanghai and Qingdao.
Here, we present a detailed introduction of the construction of the dataset, and show its two representative exemplary applications, namely SAR-optical image matching and SAR ship detection boosted by cross-modal information from optical images. As a large open SAR-optical dataset with multiple scenes of a high resolution, we believe QXS-SAROPT will be of potential value for further research in SAR-optical data fusion technology based on deep learning.

\end{abstract}

}

\maketitle

\IEEEdisplaynontitleabstractindextext

%
\IEEEpeerreviewmaketitle

\section{Introduction}

%
%
%
%

\IEEEPARstart{W}{ith} the rapid development of deep learning, remarkable breakthroughs have been made in deep learning-based land use segmentation, scene classification, object detection and recognition 
on the field of remote sensing in the past decade~\cite{zhang2016deep,zhu2017deep,2017Comprehensive,tsagkatakis2019survey}. This is mainly due to the powerful feature extraction and representation ability of deep neural networks~\cite{sainath2013deep,simonyan2014very,Schmidhuber2015Deep,he2016deep}, which can well map the remote sensing observations into the desired geographical knowledge. However, the current mainstream remote sensing image interpretation technology is still mainly focused on mono-modal data, and cannot make full use of the complementary and correlated information of multimodal data from different sensors with heterogenous characteristics, resulting in insufficient intelligent interpretation capabilities and limited application scenarios. For example, optical imaging is easily restricted by illumination and weather conditions, based on which accurate interpretation cannot be obtained at night or under complex weather with clouds, fog and so on. Compared with optical imaging, Synthetic Aperture Radar (SAR) imaging can achieve full-time and all-weather earth observations, however, it is difficult to interpret the SAR images with less texture features, even for well-trained experts. Therefore, gathering sufficient amounts of training SAR data with diverse scenes and accurate labeling is a challenging problem, which heavily affects the deep research and application of SAR image based intelligent interpretation.

To address the above issues, multimodal data fusion~\cite{Michael2016Data,zhang2018change,feng2019embranchment,zhang2019detecting} becomes one of the most promising directions of deep learning in remote sensing, especially the combined utilization of SAR and optical data because these data modalities are completely different from each other both in terms of geometric and radiometric appearance
~\cite{schmitt2017fusion,schmitt2018sen1,feng2019integrating,kulkarni2020pixel,li2020multimodal}. To promote the development of research in SAR-optical data fusion based on deep learning, it is very important to obtain large datasets of perfectly aligned images or image patches. However, collecting such a large amount of aligned multi-sensor image data is a very time-consuming and labor-intensive task~\cite{wang2019challenge}. Moreover, the existing SAR-optical patch matching dataset either lacks of scene diversity due to the huge difficulty in pixel-level matching between optical and SAR images~\cite{wang2018sarptical}, or has a low resolution limited by the remote sensing satellites used for data acquisition~\cite{schmitt2018sen1}, or covers only a single area~\cite{shermeyer2020spacenet}. 

Based on the analysis above, in this paper, we publish the so-called QXS-SAROPT dataset containing 20,000 SAR-optical patch-pairs from multiple scenes of a high resolution of 1 meter. Specifically, the patches are collected from images acquired by the SAR satellite GaoFen-3~\cite{zhang2017system} and optical satellites used for Google Earth~\cite{2017Google}. These images spread across land masses of San Diego, Shanghai and Qingdao. The QXS-SAROPT dataset under open access license CCBY is publicly available at~\url{https://github.com/yaoxu008/QXS-SAROPT}. In the rest of this paper, we will introduce the detailed construction process of the dataset, its two example applications, as well as its strengths and limitations.

\section{Construction of the QXS-SAROPT dataset}
Figure~\ref{fig:construction} shows the procedure for the QXS-SAROPT dataset construction. As shown in Figure~\ref{fig:construction}, the procedure contains the following seven steps:

\begin{figure}[!htb]
	\centering
	\includegraphics[scale=0.25]{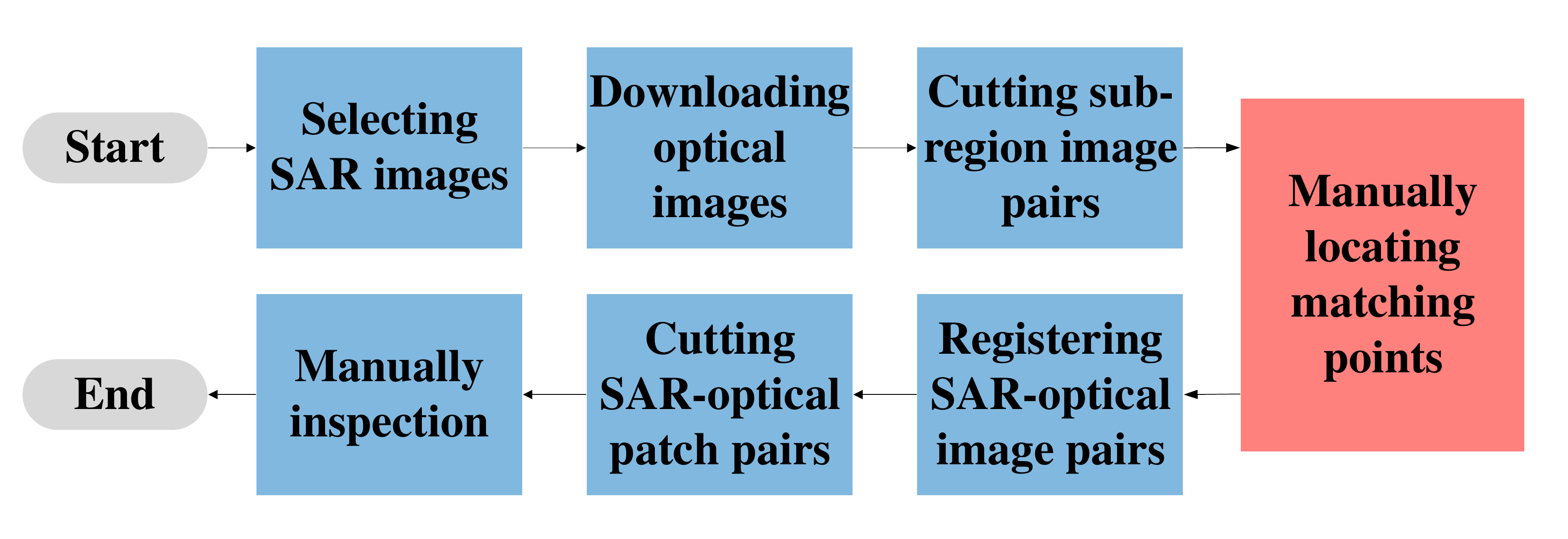}	\caption{Flowchart of the procedure for construction of the QXS-SAROPT dataset.}
	\label{fig:construction}
\end{figure}
\subsection{Selecting SAR images}
\begin{figure}[!htb]
	\centering
	\subfigure[San Diego]{
		\label{fig:subfig:a} 
	    \includegraphics[scale=0.055]{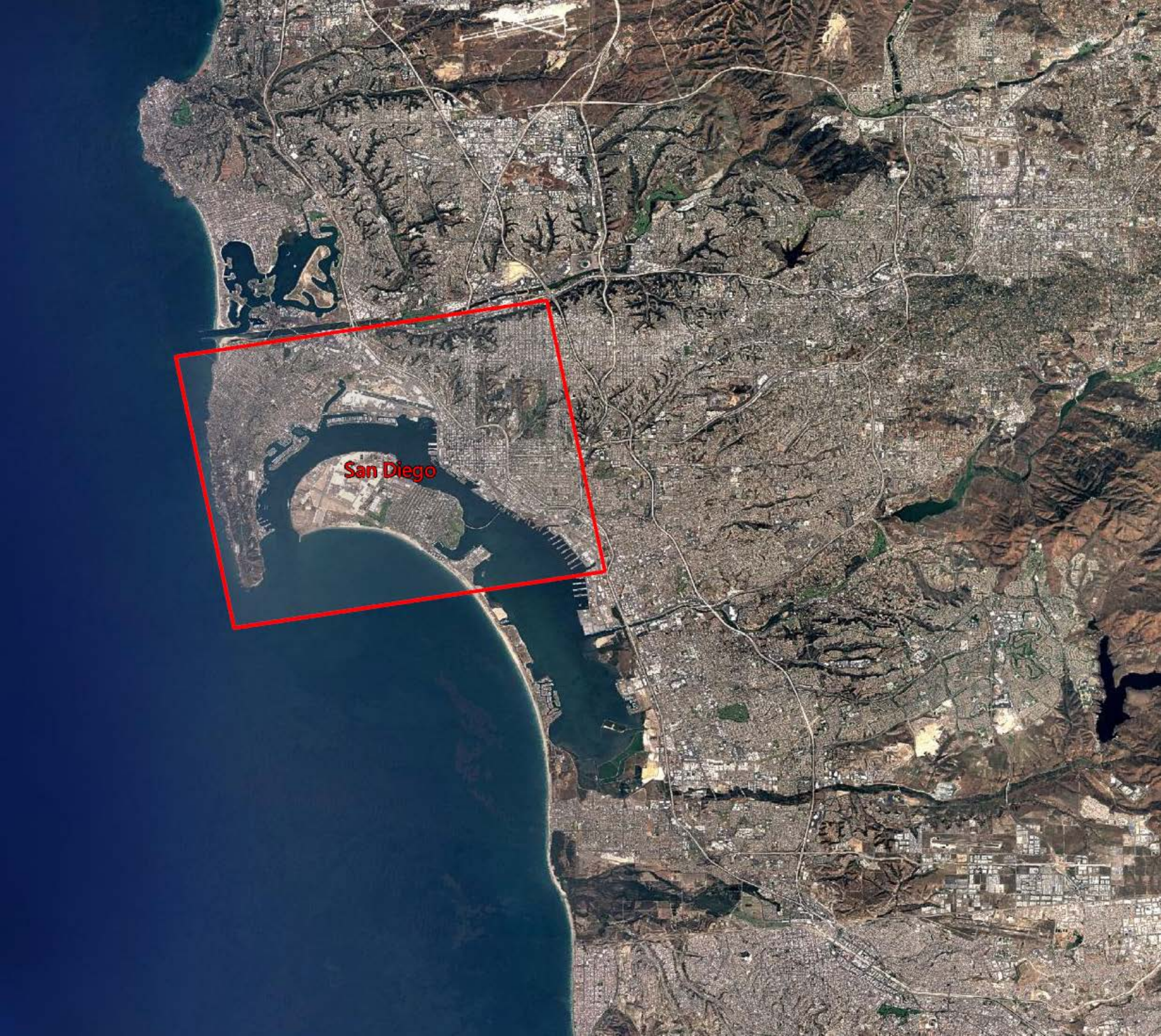}}
	\subfigure[Shanghai]{
		\label{fig:subfig:b} 
	    \includegraphics[scale=0.05]{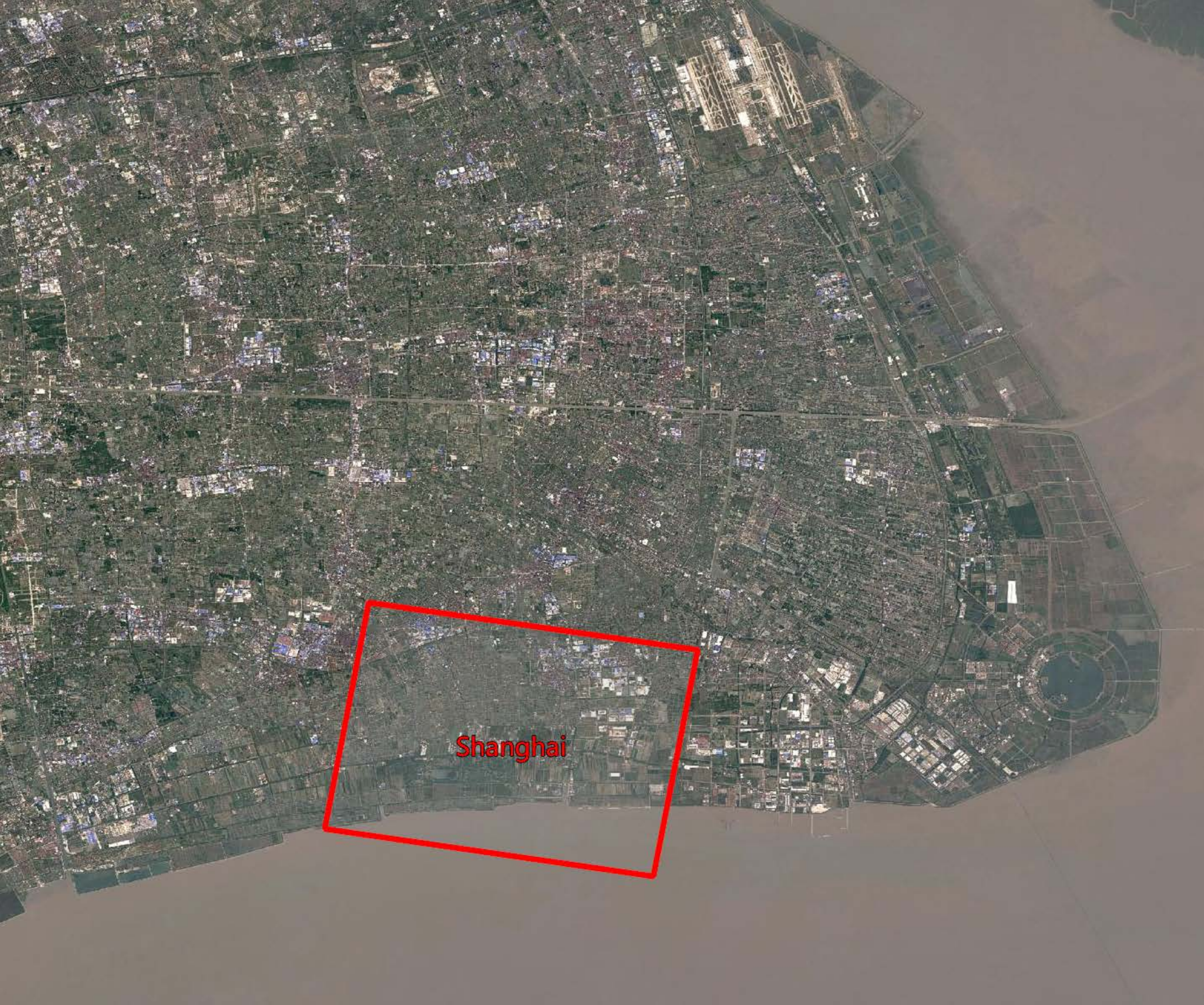}}
	\subfigure[Qingdao]{
		\label{fig:subfig:c} 
	    \includegraphics[scale=0.05]{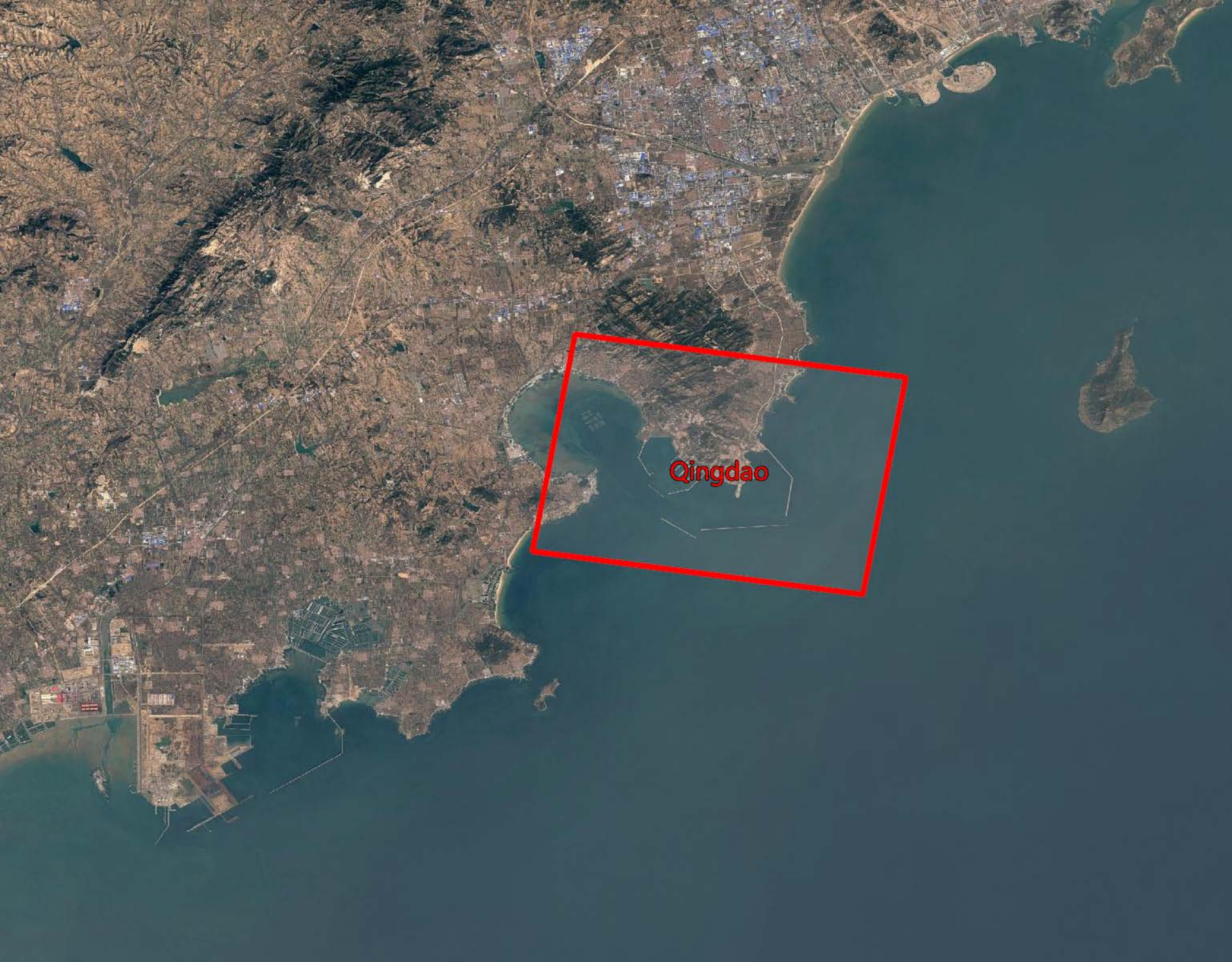}}
	\caption{Coverage of the SAR images of Gaofen-3 for constructing the QXS-SAROPT dataset. The red rectangles indicate the coverage of each image.}
	\label{fig:coverage}
\end{figure}

For application scenarios such as land cover segmentation, scene classification, detection and recognition, we first select three SAR images acquired by the Gaofen-3 satellite~\cite{zhang2017system} that contain rich land cover types such as inland, offshore, and mountains. Gaofen-3, developed by China Aerospace Science and Technology Corporation, is a part of the Chinese Gaofen (High-Resolution) Earth Observation Project. It is the first C-band and multi-polarization SAR imaging satellite in low earth orbit with a resolution of 1 meter in China~\cite{sun2017The}. Our SAR data originates from the spotlight mode images with single polarization. The spatial resolution of SAR imagery is $1m \times 1m$ each pixel. The area of each image is 100 square kilometers from three big port cities: San Diego, Shanghai, and Qingdao. Image sizes are 14624 $\times$ 33820, 17080 $\times$ 28778, and 17080 $\times$ 28946 respectively. Details of these images, including resolution, swath, incidence angle, and polarization are presented in Table~\ref{table:information}. The coverage of these images is shown in Figure~\ref{fig:coverage}.

\renewcommand{\arraystretch}{1.5} 
\begin{table*}[!htb]
	\centering
	\caption{Detailed information of the SAR images of Gaofen-3 for constructing the QXS-SAROPT dataset.}
	\label{table:information}
	\setlength{\tabcolsep}{3.5mm}{
		\begin{tabular}{|c|c|c|c|c|c|c|c|}	
			\hline
			No & Coverage region & Imaging mode & \makecell*[c]{Resolution\\Rg.$\times$Az.(m)} &  \makecell*[c]{Swath\\(km)} & \makecell*[c]{Incident Angle\\($^{\circ}$)} & Polarization 
			& Image size \\ 
			\hline
			1 & San Diego & \multirow{3}{*}{spotlight} & \multirow{3}{*}{1$\times$1} & \multirow{3}{*}{10} & \multirow{3}{*}{20\textasciitilde50} & \multirow{3}{*}{single} & 14624$\times$33820 \\
			\cline{1-2} \cline{8-8}
			2 & Shanghai &  &  &  &  &  & 17080$\times$28778 \\ 
			\cline{1-2} \cline{8-8}
		    3 & Qingdao &  &  &  &  &  & 17080$\times$28946 \\
			\hline
	\end{tabular}}

\end{table*}

\subsection{Downloading corresponding optical images}
Using the virtual globe software Google Earth~\cite{2017Google}, which maps the earth by superimposing satellite images, aerial photography, and GIS data onto a 3D globe, we download the optical images of the corresponding area with a resolution of $1m$ provided by Maxar Technologies and INEGI. Given the latitude and longitude coordinates of the desired area and the specified resolution, the corresponding optical images are selected and downloaded, where the image patches are extracted.

\subsection{Cutting SAR-optical image pairs into sub-region image pairs}
Since SAR follows the electromagnetic imaging mechanism, which would cause geometric distortion, such as foreshortening, layover and radar shadow~\cite{hughes2020deep}, resulting in obvious differences between SAR images and optical images in appearance. Therefore, it is almost impossible to accurately register the whole optical image with the whole corresponding SAR image~\cite{xiang2019flow}. Taking the above issues into consideration, we cut the whole SAR-optical image pair into several sub-region image pairs according to the complexity of land coverage. After that, we can register the sub-region image pairs separately instead of directly registering the whole image pair.

\subsection{Manually locating matching points of sub-region SAR-optical image pairs}
\begin{figure}[!htb]
		\centering
		\subfigure[Optical]{
			\label{fig:matchingpoint:a} 
            \includegraphics[scale=0.138]{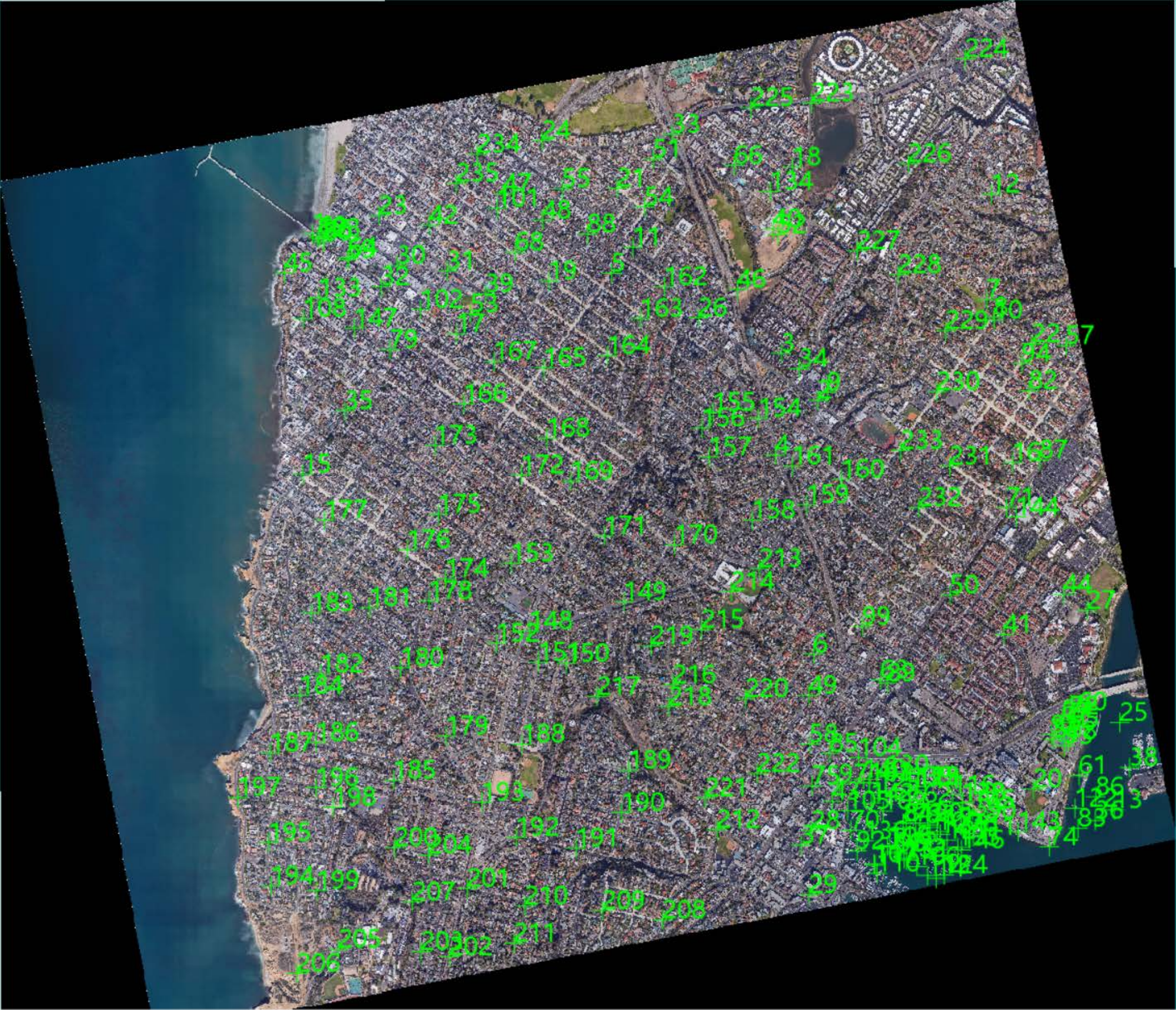}}
		\subfigure[SAR]{
			\label{fig:matchingpoint:b} 
            \includegraphics[scale=0.138]{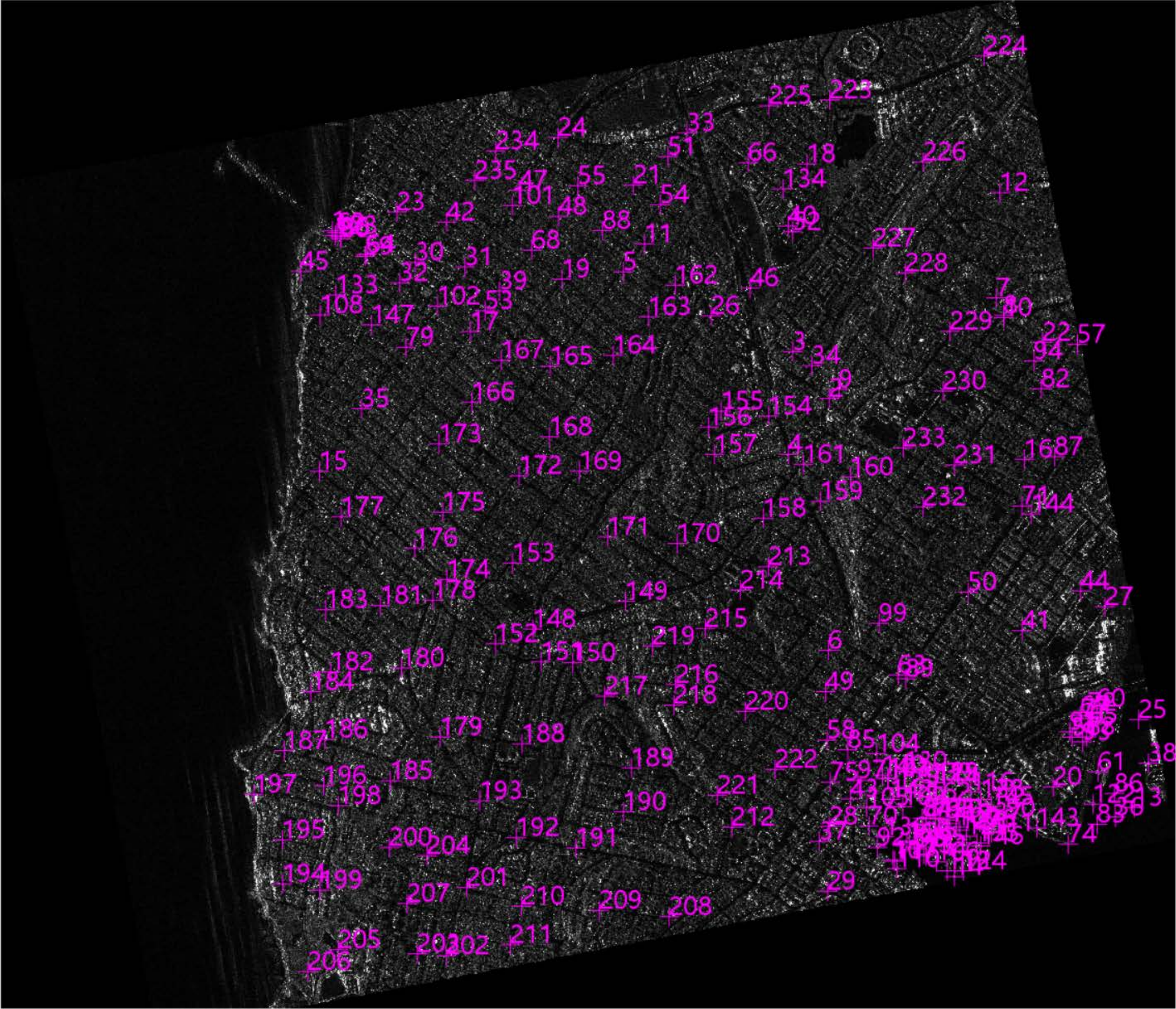}}
	\caption{Illustration of matching points manually selected for one exemplary sub-region SAR-optical image pair.}
	\label{fig:matchingpoint}
\end{figure}

As mentioned above, due to different imaging mechanism, optical and SAR imagery are completely differ from each other both in terms of geometric and radiometric appearance. Therefore, registering them directly based on matching points located automatically may lead to a bad performance since the existing image registration approaches~\cite{Zitova03imageregistration} are mainly designed for registering optical images. In order to improve the registration performance of these two modalities, matching points of the sub-region SAR-optical image pairs are manually located, which are selected as the geometrically invariant corner points of buildings, ships, roads, etc. Figure~\ref{fig:matchingpoint} shows matching points manually selected for one exemplary sub-region SAR-optical image pair. 

\subsection{Registering sub-region SAR-optical image pairs}
\begin{figure}[!htb]
	\centering
	\subfigure[Optical]{
	    \label{fig:registration:a} 
	    \includegraphics[scale=0.13]{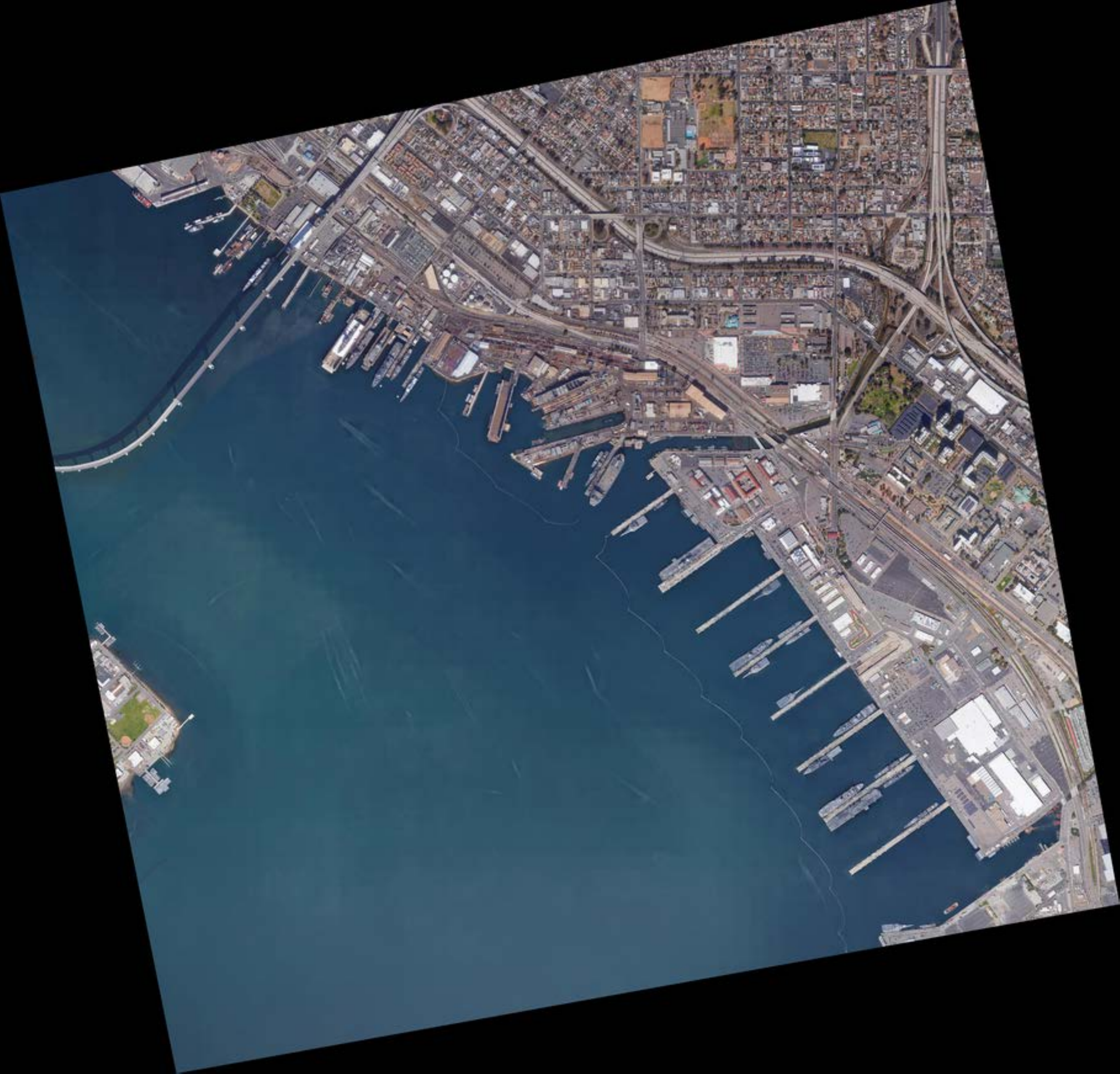}}
	\subfigure[SAR]{
	    \label{fig:registration:b} 
	    \includegraphics[scale=0.13]{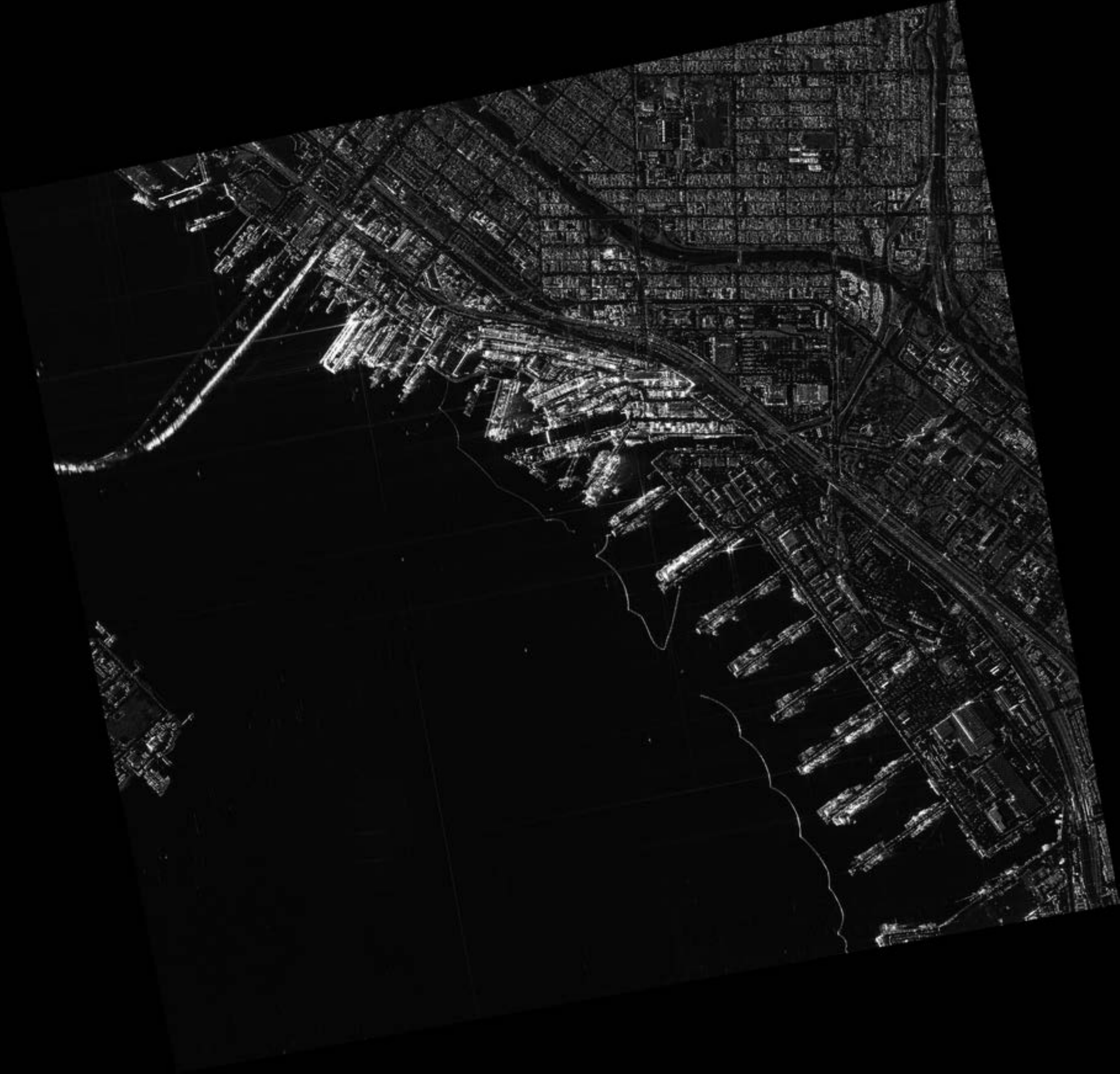}}
	\caption{Illustration of an exemplary registered sub-region SAR-optical image pair.}
	\label{fig:registration}
\end{figure}

With the manually located matching points, we use an existing automatic image registration software to register the sub-region SAR-optical image pairs. Optical imagery is registered to the fixed SAR image through the bilinear interpolation method. Figure~\ref{fig:registration} shows an example of registered sub-region SAR-optical image pair.

\subsection{Cutting registered sub-region SAR-optical image pairs into patch-pairs} 
Since the dataset we build is intended for deep learning, the registered sub-region SAR-optical image pairs are supposed to be cropped into small patches of 256 $\times$ 256 pixels to fit the neural network. Aiming at maximizing the number of patches from the available scenes as well as reducing the overlap between adjacent patches, we cut the image pairs with a stride of 52 to make the 20\% overlap between nearby patches. After completing this step, we obtain 46071 SAR-optical patch pairs.

\subsection{Manual inspection}
\begin{figure}[!htb]
	\centering
    \includegraphics[scale=0.4]{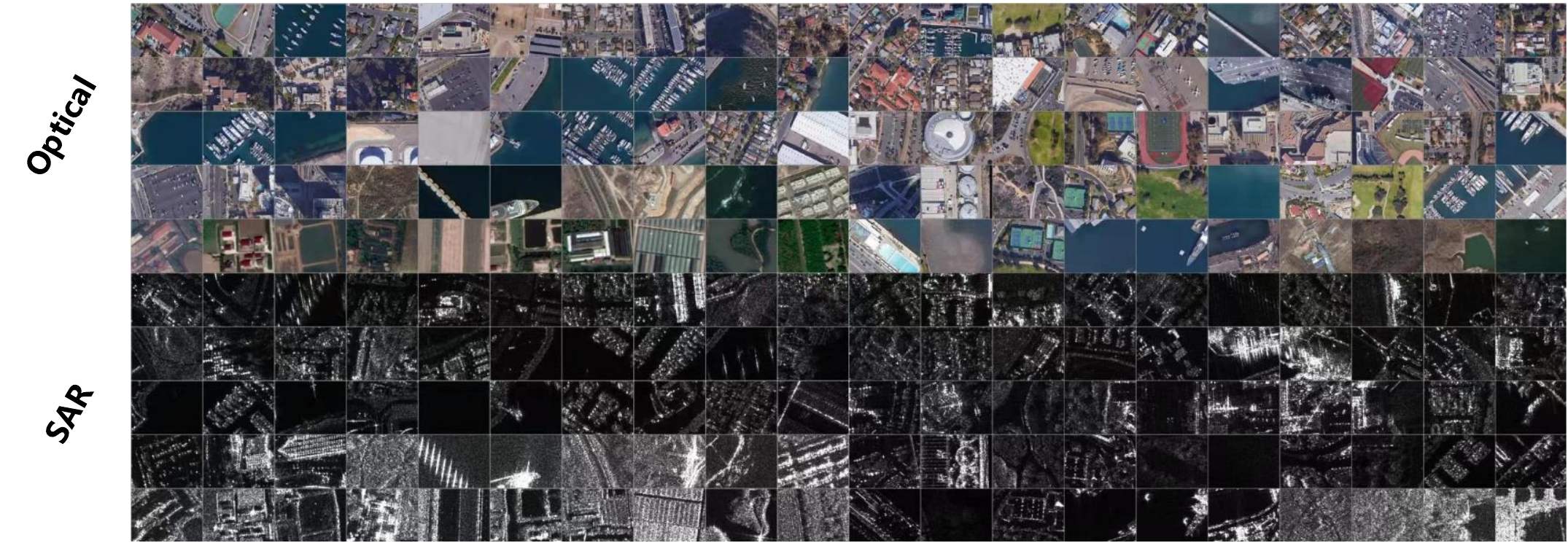}
	\caption{Some exemplary patch-pairs from the QXS-SAROPT dataset. Top part: Google Earth RGB image patches, bottom part: GaoFen-3 SAR image patches.}
	\label{fig:patchpairs}
\end{figure}

At last, we have double-checked all patches manually to ensure that every image contains meaningful information and texture. Therefore, we remove indistinguishable or flawed images, such as images with similar scenes, texture-less sea or visible mosaicking seamlines. Finally, 20,000 high-quality image pairs are preserved, some of which are shown in Figure~\ref{fig:patchpairs} for examples.
 
\section{Example applications}
Two example applications on this dataset are introduced in this section: SAR-optical image matching and SAR ship detection, aiming to stimulate the idea of using this dataset and guide further exploration and research of SAR-optical data fusion based on deep learning.

\subsection{SAR-optical image matching}
\begin{figure}[!htb]
	\centerline{\includegraphics[scale=0.29]{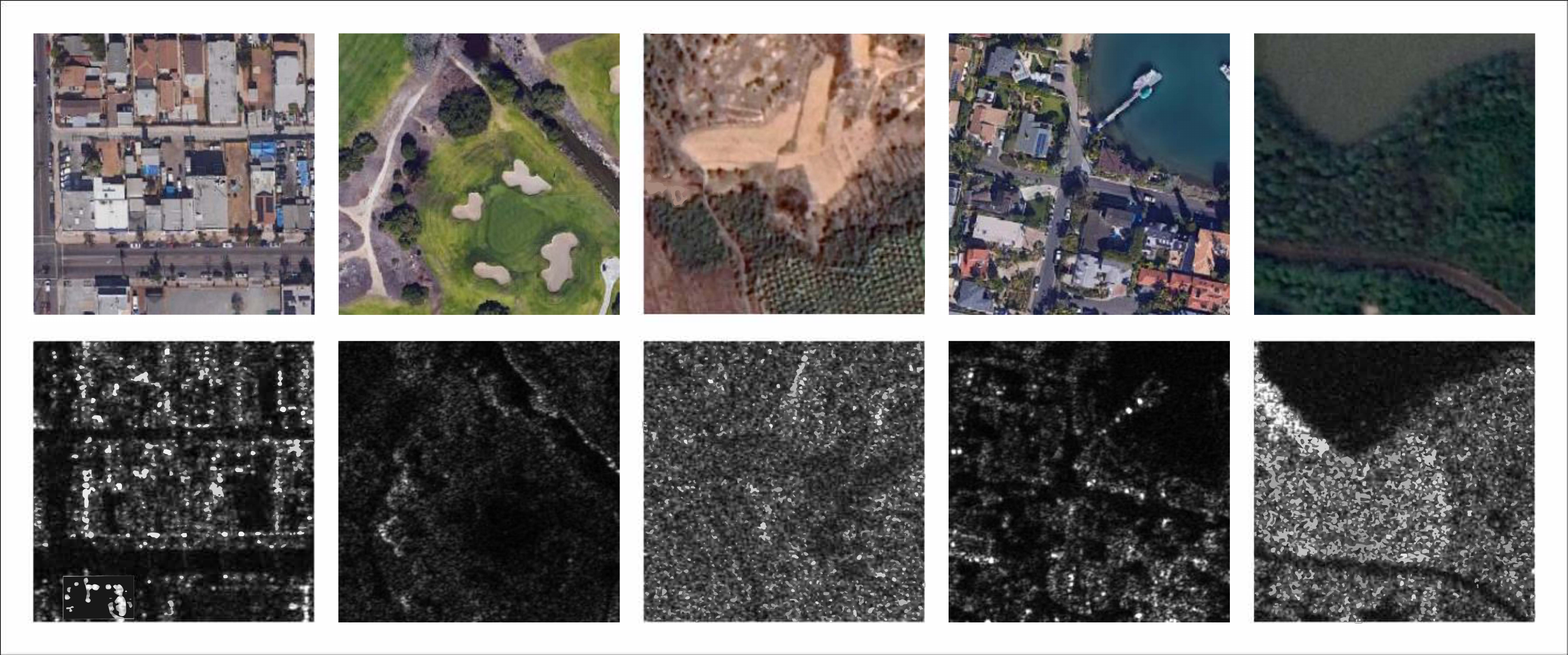}}
	\caption{Some correctly identified matching patch-pairs by BNN~\cite{xu2019task} for a test subset of QXS-SAROPT. Top row: optical images, bottom row: corresponding SAR images.}
	\label{fig:matchpatches}
\end{figure}

Multi-modal image matching still remains a challenge for the existing methods mainly striving to measure similarity between mono-modal imagery. Specifically, matching SAR-optical images is extremely difficult due to the vast geometric and radiometric differences~\cite{mou2017cnn,hughes2018identifying,hughes2020deep,xiong2020deep,zhang2018retrieval}. The QXS-SAROPT dataset can help in SAR-optical image matching by providing a large amount of training data for the deep learning methods, such as that proposed in~\cite{xu2019task}. This method uses a bridge neural network (BNN) architecture to project the corresponding SAR-optical image patches of the QXS-SAROPT dataset into a common feature subspace, where similarities between the embedding features can be easily measured. The matching accuracy of a test subset can reach 82.9\% and 82.8\% with the model of~\cite{bao2021boosting} trained on 70\% patch-pairs of the QXS-SAROPT dataset using the ResNet50~\cite{he2016deep} and Darknet53~\cite{redmon2018yolov3} backbone, respectively. The detailed results can be seen in Table~\ref{tab:bnn}, and some exemplary matches correctly identified for the test subset are shown in Figure~\ref{fig:matchpatches}.
\renewcommand{\arraystretch}{1.1} 
\begin{table}[!htb]  
	\centering
	\setlength{\tabcolsep}{4mm}   
	\begin{threeparttable} 
		\caption{Results for BNN~\cite{xu2019task} patch-matching trained on QXS-SAROPT.}
		\begin{tabular}{cccc}
			\toprule
			Backbone&Accuracy&Precision&Recall \\
			\midrule  
			ResNet50~\cite{he2016deep}&$0.829$&$0.748$&$0.993$ \cr 
			Darknet53~\cite{redmon2018yolov3}&$0.828$&$0.746$&$0.995$ \cr
			\bottomrule
		\end{tabular} 
		\label{tab:bnn} 
	\end{threeparttable}  
\end{table}

\subsection{SAR ship detection}
\begin{figure}[!htb]
	\centerline{\includegraphics[scale=0.3]{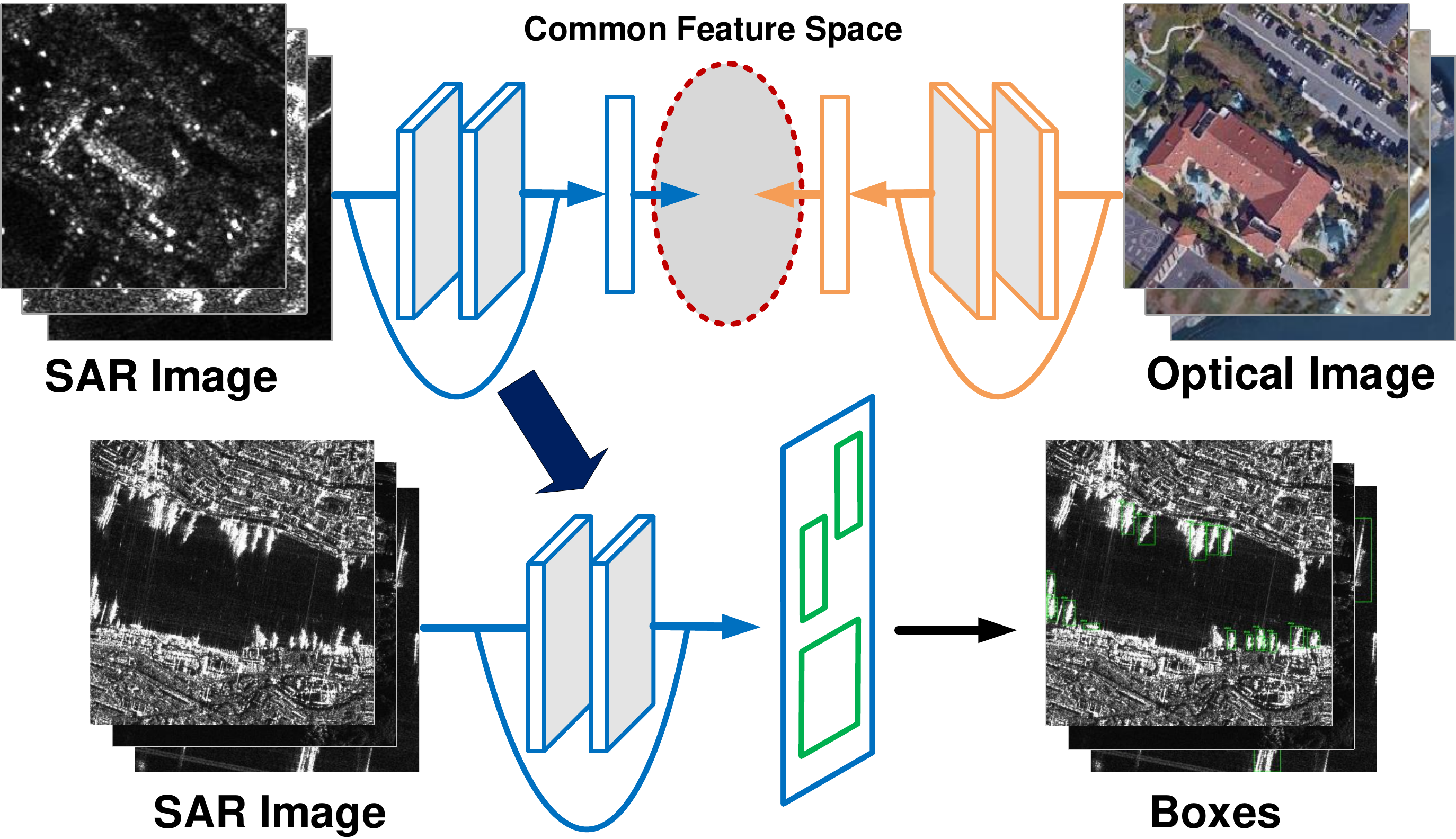}}
	\caption{The overall process to implement SAR ship detection using BNN~\cite{xu2019task} based on the QXS-SAROPT dataset.}
	\label{fig:crl}
\end{figure}

SAR ship detection in complex scenes is a great challenging task. Because of powerful feature embedding ability, convolutional neural networks (CNN)~\cite{sainath2013deep}-based SAR ship detection methods have drawn considerable attention. The pretraining technique is usually adopted to support these CNN-based SAR ship detectors due to the scarce labeled SAR images. However directly leveraging ImageNet~\cite{Russakovsky2015ImageNet} pretraining is hardly to obtain a good ship detector because of different imaging geometry between optical and SAR images. The QXS-SAROPT dataset can provide a platform for SAR ship detection, such as~\cite{bao2021boosting,competitation2020} proposing an optical-SAR matching (OSM) pretraining technique to enhance the general feature embedding of SAR images by BNN~\cite{xu2019task} based on the QXS-SAROPT dataset. BNN can transfer plentiful texture features from optical images to SAR images and the SAR CNN can be further used as the backbone of the detection framework to perform SAR ship detection. The overall process to implement SAR ship detection using BNN based on the QXS-SAROPT dataset is depicted in~\ref{fig:crl}. As shown in Table~\ref{tab:osm-ssd}, compared with ImageNet pretraining based SAR ship detector (ImageNet-SSD), the Average Precision (AP) of detection results of OSM pretraining based SAR ship detector (OSM-SSD) on SAR ship detection dataset AIR-SARShip-1.0~\cite{xian2019air} can be improved by 1.32\% and 1.24\% using two-stage detection benchmark: Faster R-CNN~\cite{ren2015faster} and one-stage detection benchmark: YOLOv3~\cite{redmon2018yolov3}, respectively. For more detailed results, please refer to~\cite{bao2021boosting,competitation2020}.

\renewcommand{\arraystretch}{1.1} 
\begin{table}[!htb]  
	\centering
	\setlength{\tabcolsep}{3mm}   
	\begin{threeparttable} 
		\caption{SAR ship detection results on AIR-SARShip-1.0~\cite{xian2019air} using ImageNet-SSD pretrained on ImageNet and OSM-SSD~\cite{bao2021boosting,competitation2020} pretrained on QXS-SAROPT.}
		\begin{tabular}{ccc}
			\toprule
			&Faster R-CNN~\cite{ren2015faster}& YOLOv3~\cite{redmon2018yolov3} \\
			\midrule  
			ImageNet-SSD& $0.8720$ &  $0.8712$ \cr 
			OSM-SSD& $0.8852$ & $0.8836$ \cr
			\bottomrule
		\end{tabular} 
		\label{tab:osm-ssd} 
	\end{threeparttable}  
\end{table}

\section{Strengths and limitations of the dataset}
As far as we know, QXS-SAROPT is the first dataset that provides high-resolution ($1m$) co-registered SAR and optical satellite image patches covering over three big port cities in the world. The other three existing datasets in this field are the SARptical dataset~\cite{wang2018sarptical}, SEN1-2 dataset~\cite{schmitt2018sen1} and SpaceNet 6 dataset~\cite{shermeyer2020spacenet}.

SARptical provides high-resolution image patches from TerraSAR-X and aerial hotogrammetry, but it only extracts 10,000 patches from inland area focusing on urban buildings, and many extracted patches contain more than 50\% overlap while the whole size of each patch is only 112$\times$112, which may be difficult to satisfy the requirment of deep learning for large amounts of data. In contrast, QXS-SAROPT has a boarder coverage and extracts 20,000 patch-pairs from multiple scenes, e.g. inland, sea, mountain area and so forth. QXS-SAROPT focuses not only on buildings but also harbours and a variety of land types. Therefore, it will be a valuable dataset for many researches in SAR-optical data fusion and deep learning for remote sensing, such as scene classification. 

Regarding SEN1-2, it is the largest dataset of SAR-optical images with 282,384 pairs of corresponding image patches spread over the world and all meteorological seasons. However, it is built on Sentinel-2 satellite with a low resolution of $5m$, which may be not applicable for learning of small-sized objects, such as ships. Conversely, QXS-SAROPT can well enhance the learning ability of small-sized objects because of the high resolution of GaoFen-3.

As for SpaceNet 6, whose spatial resolution comes to $0.5m$ per pixel with SAR images provided by Capella Space and optical imagery provided by the Maxar
Worldview-2 satellite. This dataset concentrates on the largest port in Europe: Rotterdam, the Netherlands. It provides an effective platform for SAR-optical fusion task due to the covery for different kinds of objects and land types. However, only one city covered in this dataset. Different from SpaceNet 6, QXS-SAROPT tries to cover more land area with SAR images provided by the Gaofen-3 satellite. 

As mentioned above, QXS-SAROPT is expected to benefit the multi-modal data fusion task a lot. However, QXS-SAROPT also has its limitations: it only covers three port cities now, which is still limited in both coverage and scenes. Furthermore, fixing the patch size as 256 $\times$ 256 may be inadaptable to different tasks. A future version 2 will be released to extend the dataset accordingly.

\section{Summary and conclusion}
In this paper, we have introduced and published the QXS-SAROPT dataset, aiming at promoting the development of SAR-optical data fusion of satellite remote sensing based on deep learning. The dataset contains 20,000 pairs of SAR and optical image patches with a high resolution of 1m extracted from multiple Gaofen-3 and Google Earth scenes. To further explore the potential value of the dataset, We are going to release a improved version of the dataset in future, which will cover more land areas with versatile scenes and has different sized patch-pairs suitable for various SAR-optical data fusion tasks. Moreover, for benefitting the research in object detection and recognition fusing SAR-optical data, position and label annotations might be added for objects of interest to every patch-pair of the dataset.

\ifCLASSOPTIONcaptionsoff
  \newpage
\fi



\bibliographystyle{IEEEtran}
\bibliography{egbib}

\begin{thebibliography}{10}
\providecommand{\url}[1]{#1}
\csname url@samestyle\endcsname
\providecommand{\newblock}{\relax}
\providecommand{\bibinfo}[2]{#2}
\providecommand{\BIBentrySTDinterwordspacing}{\spaceskip=0pt\relax}
\providecommand{\BIBentryALTinterwordstretchfactor}{4}
\providecommand{\BIBentryALTinterwordspacing}{\spaceskip=\fontdimen2\font plus
\BIBentryALTinterwordstretchfactor\fontdimen3\font minus
  \fontdimen4\font\relax}
\providecommand{\BIBforeignlanguage}[2]{{%
\expandafter\ifx\csname l@#1\endcsname\relax
\typeout{** WARNING: IEEEtran.bst: No hyphenation pattern has been}%
\typeout{** loaded for the language `#1'. Using the pattern for}%
\typeout{** the default language instead.}%
\else
\language=\csname l@#1\endcsname
\fi
#2}}
\providecommand{\BIBdecl}{\relax}
\BIBdecl

\bibitem{zhang2016deep}
L.~{Zhang}, L.~{Zhang}, and B.~{Du}, ``Deep learning for remote sensing data: A
  technical tutorial on the state of the art,'' \emph{IEEE Geoscience and
  Remote Sensing Magazine}, vol.~4, no.~2, pp. 22--40, 2016.

\bibitem{zhu2017deep}
X.~X. Zhu, D.~Tuia, L.~Mou, G.-S. Xia, L.~Zhang, F.~Xu, and F.~Fraundorfer,
  ``Deep learning in remote sensing: A comprehensive review and list of
  resources,'' \emph{IEEE Geoscience and Remote Sensing Magazine}, vol.~5,
  no.~4, pp. 8--36, 2017.

\bibitem{2017Comprehensive}
J.~E. Ball, D.~T. Anderson, and C.~S. Chan, ``Comprehensive survey of deep
  learning in remote sensing: theories, tools, and challenges for the
  community,'' \emph{Journal of Applied Remote Sensing}, vol.~11, no.~4, 2017.

\bibitem{tsagkatakis2019survey}
G.~Tsagkatakis, A.~Aidini, K.~Fotiadou, M.~Giannopoulos, A.~Pentari, and
  P.~Tsakalides, ``Survey of deep-learning approaches for remote sensing
  observation enhancement,'' \emph{Sensors}, vol.~19, no.~18, p. 3929, 2019.

\bibitem{sainath2013deep}
T.~N. Sainath, A.-r. Mohamed, B.~Kingsbury, and B.~Ramabhadran, ``Deep
  convolutional neural networks for lvcsr,'' in \emph{2013 IEEE international
  conference on acoustics, speech and signal processing}.\hskip 1em plus 0.5em
  minus 0.4em\relax IEEE, 2013, pp. 8614--8618.

\bibitem{simonyan2014very}
K.~Simonyan and A.~Zisserman, ``Very deep convolutional networks for
  large-scale image recognition,'' \emph{arXiv preprint arXiv:1409.1556}, 2014.

\bibitem{Schmidhuber2015Deep}
Schmidhuber and Jürgen, ``Deep learning in neural networks: An overview,''
  \emph{Neural Netw}, vol.~61, pp. 85--117, 2015.

\bibitem{he2016deep}
K.~He, X.~Zhang, S.~Ren, and J.~Sun, ``Deep residual learning for image
  recognition,'' in \emph{Proc. CVPR}, 2016, pp. 770--778.

\bibitem{Michael2016Data}
Michael, Schmitt, Xiao, Xiang, and Zhu, ``Data fusion and remote sensing: An
  ever-growing relationship,'' \emph{IEEE Geoscience and Remote Sensing
  Magazine}, vol.~4, no.~4, pp. 6--23, 2016.

\bibitem{zhang2018change}
Z.~Zhang, G.~Vosselman, M.~Gerke, D.~Tuia, and M.~Y. Yang, ``Change detection
  between multimodal remote sensing data using siamese cnn,'' \emph{arXiv
  preprint arXiv:1807.09562}, 2018.

\bibitem{feng2019embranchment}
P.~Feng, Y.~Lin, J.~Guan, Y.~Dong, G.~He, Z.~Xia, and H.~Shi, ``Embranchment
  cnn based local climate zone classification using sar and multispectral
  remote sensing data,'' in \emph{IGARSS 2019-2019 IEEE International
  Geoscience and Remote Sensing Symposium}.\hskip 1em plus 0.5em minus
  0.4em\relax IEEE, 2019, pp. 6344--6347.

\bibitem{zhang2019detecting}
Z.~Zhang, G.~Vosselman, M.~Gerke, C.~Persello, D.~Tuia, and M.~Y. Yang,
  ``Detecting building changes between airborne laser scanning and
  photogrammetric data,'' \emph{Remote sensing}, vol.~11, no.~20, p. 2417,
  2019.

\bibitem{schmitt2017fusion}
M.~Schmitt, F.~Tupin, and X.~X. Zhu, ``Fusion of sar and optical remote sensing
  data--challenges and recent trends,'' in \emph{2017 IEEE International
  Geoscience and Remote Sensing Symposium (IGARSS)}.\hskip 1em plus 0.5em minus
  0.4em\relax IEEE, 2017, pp. 5458--5461.

\bibitem{schmitt2018sen1}
M.~Schmitt, L.~H. Hughes, and X.~X. Zhu, ``The sen1-2 dataset for deep learning
  in sar-optical data fusion,'' \emph{arXiv preprint arXiv:1807.01569}, 2018.

\bibitem{feng2019integrating}
Q.~Feng, J.~Yang, D.~Zhu, J.~Liu, H.~Guo, B.~Bayartungalag, and B.~Li,
  ``Integrating multitemporal sentinel-1/2 data for coastal land cover
  classification using a multibranch convolutional neural network: A case of
  the yellow river delta,'' \emph{Remote Sensing}, vol.~11, no.~9, p. 1006,
  2019.

\bibitem{kulkarni2020pixel}
S.~C. Kulkarni and P.~P. Rege, ``Pixel level fusion techniques for sar and
  optical images: A review,'' \emph{Information Fusion}, vol.~59, pp. 13--29,
  2020.

\bibitem{li2020multimodal}
X.~Li, L.~Lei, Y.~Sun, M.~Li, and G.~Kuang, ``Multimodal bilinear fusion
  network with second-order attention-based channel selection for land cover
  classification,'' \emph{IEEE Journal of Selected Topics in Applied Earth
  Observations and Remote Sensing}, vol.~13, pp. 1011--1026, 2020.

\bibitem{wang2019challenge}
Y.~Wang and X.~X. Zhu, ``The challenge of creating the sarptical dataset,'' in
  \emph{IGARSS 2019-2019 IEEE International Geoscience and Remote Sensing
  Symposium}.\hskip 1em plus 0.5em minus 0.4em\relax IEEE, 2019, pp.
  5714--5717.

\bibitem{wang2018sarptical}
------, ``The sarptical dataset for joint analysis of sar and optical image in
  dense urban area,'' in \emph{IGARSS 2018-2018 IEEE International Geoscience
  and Remote Sensing Symposium}.\hskip 1em plus 0.5em minus 0.4em\relax IEEE,
  2018, pp. 6840--6843.

\bibitem{shermeyer2020spacenet}
J.~Shermeyer, D.~Hogan, J.~Brown, A.~Van~Etten, N.~Weir, F.~Pacifici,
  R.~Hansch, A.~Bastidas, S.~Soenen, T.~Bacastow \emph{et~al.}, ``Spacenet 6:
  Multi-sensor all weather mapping dataset,'' in \emph{Proceedings of the
  IEEE/CVF Conference on Computer Vision and Pattern Recognition Workshops},
  2020, pp. 196--197.

\bibitem{zhang2017system}
Q.~Zhang, ``System design and key technologies of the gf-3 satellite,''
  \emph{Acta Geodaetica et Cartographica Sinica}, vol.~46, no.~3, pp. 269--277,
  6 2017.

\bibitem{2017Google}
\url{https://earth.google.com/},.

\bibitem{sun2017The}
J.~Sun, W.~Yu, and Y.~Deng, ``The sar payload design and performance for the
  gf-3 mission,'' \emph{Sensors}, vol.~17, no.~10, p. 2419, 2017.

\bibitem{hughes2020deep}
L.~H. Hughes, D.~Marcos, S.~Lobry, D.~Tuia, and M.~Schmitt, ``A deep learning
  framework for matching of sar and optical imagery,'' \emph{ISPRS Journal of
  Photogrammetry and Remote Sensing}, vol. 169, pp. 166--179, 2020.

\bibitem{xiang2019flow}
Y.~Xiang, F.~Wang, L.~Wan, N.~Jiao, and H.~You, ``Os-flow: A robust algorithm
  for dense optical and sar image registration,'' \emph{IEEE Transactions on
  Geoscience and Remote Sensing}, vol.~57, no.~9, pp. 6335--6354, 2019.

\bibitem{Zitova03imageregistration}
B.~Zitová and J.~Flusser, ``Image registration methods: a survey,''
  \emph{Image and Vision Computing}, vol.~21, pp. 977--1000, 2003.

\bibitem{mou2017cnn}
L.~Mou, M.~Schmitt, Y.~Wang, and X.~X. Zhu, ``A cnn for the identification of
  corresponding patches in sar and optical imagery of urban scenes,'' in
  \emph{2017 Joint Urban Remote Sensing Event (JURSE)}.\hskip 1em plus 0.5em
  minus 0.4em\relax IEEE, 2017, pp. 1--4.

\bibitem{hughes2018identifying}
L.~H. Hughes, M.~Schmitt, L.~Mou, Y.~Wang, and X.~X. Zhu, ``Identifying
  corresponding patches in sar and optical images with a pseudo-siamese cnn,''
  \emph{IEEE Geoscience and Remote Sensing Letters}, vol.~15, no.~5, pp.
  784--788, 2018.

\bibitem{xiong2020deep}
W.~Xiong, Z.~Xiong, Y.~Zhang, Y.~Cui, and X.~Gu, ``A deep cross-modality
  hashing network for sar and optical remote sensing images retrieval,''
  \emph{IEEE Journal of Selected Topics in Applied Earth Observations and
  Remote Sensing}, vol.~13, pp. 5284--5296, 2020.

\bibitem{zhang2018retrieval}
Y.~Zhang, W.~Zhou, and H.~Li, ``Retrieval across optical and sar images with
  deep neural network,'' in \emph{Pacific Rim Conference on Multimedia}.\hskip
  1em plus 0.5em minus 0.4em\relax Springer, 2018, pp. 392--402.

\bibitem{xu2019task}
Y.~Xu, X.~Xiang, and M.~Huang, ``Task-driven common representation learning via
  bridge neural network,'' in \emph{Proceedings of the AAAI Conference on
  Artificial Intelligence}, vol.~33, 2019, pp. 5573--5580.

\bibitem{bao2021boosting}
W.~Bao, M.~Huang, Y.~Zhang, Y.~Xu, X.~Liu, and X.~Xiang, ``Boosting ship
  detection in sar images with complementary pretraining techniques,''
  \emph{arXiv preprint arXiv:2103.08251}, 2021.

\bibitem{redmon2018yolov3}
J.~Redmon and A.~Farhadi, ``Yolov3: An incremental improvement,'' \emph{arXiv
  preprint arXiv:1804.02767}, 2018.

\bibitem{Russakovsky2015ImageNet}
O.~Russakovsky, J.~Deng, H.~Su, J.~Krause, S.~Satheesh, S.~Ma, Z.~Huang,
  A.~Karpathy, A.~Khosla, M.~Bernstein, A.~C. Berg, and F.-F. Li, ``Imagenet
  large scale visual recognition challenge,'' \emph{Proc. IJCV}, vol. 115,
  no.~3, pp. 211--252, 2015.

\bibitem{competitation2020}
2020 Gaofen Challenge on Automated High-Resolution Earth Observation Image
  Interpretation, online:\url{http://en.sw.chreos.org}.

\bibitem{xian2019air}
S.~Xian, W.~Zhirui, S.~Yuanrui, D.~Wenhui, Z.~Yue, and F.~Kun,
  ``Air-sarship--1.0: High resolution sar ship detection dataset,'' \emph{J.
  Radars}, vol.~8, no.~6, pp. 852--862, 2019.

\bibitem{ren2015faster}
S.~Ren, K.~He, R.~Girshick, and J.~Sun, ``Faster r-cnn: Towards real-time
  object detection with region proposal networks,'' in \emph{Proc. NIPS}, 2015,
  pp. 91--99.

\end{thebibliography}
\end{document}